\documentclass[12pt]{article}
\usepackage{mheckII}

\usepackage{manfnt}
\usepackage{longtable}

\usepackage{blkarray}

\usepackage{tikz} 
\usetikzlibrary{matrix}

\theoremstyle{theorem}

\theoremstyle{definition}

\def\Dsl{\,\raise.15ex\hbox{/}\mkern-13.5mu D}
\def\dsl{\,\raise.25ex\hbox{/}\mkern-10.5mu \partial}

\date{December, 2019}

\title{Non-minimal quasi-holes and topological degeneracy of FQHE }

\authors{ Riccardo Bergamin\footnote{e-mail: {\tt rbergamin@sissa.it}} \vskip 9pt

\centerline{SISSA, via Bonomea 265, I-34100 Trieste, ITALY}                      }

\abstract{ We extend the analysis of the Vafa $\mathcal{N}=4$ SUSY model of FQHE made in \cite{paper} and discuss other observables which characterize the FQHE topological order. We consider in particular the braiding properties of quasi-holes with generic charge. Any quasi-hole is associated with an irreducible representation of $SU(2)$ and the statistics of a bunch of $d$ punctures is described by the Knizhnik-Zamolodchikov connection specialized to the corresponding representation. We also discuss the higher genus generalization of the Vafa model of FQHE and the corresponding $tt^{*}$ geometry.}

\begin{document}

\maketitle
\newpage

\tableofcontents

\vskip40pt

\section{Introduction}

A system of interacting electrons moving on a $2d$ surface $\Sigma$ and coupled to a strong magnetic field $B$ generates at very low temperature a class of peculiar quantum phases of matter described by the fractional quantum Hall effect (FQHE) (see \cite{wen,tong} for a review). These quantum phases are classified by a 
rational number $\nu\in\bQ_{>0}$, called filling fraction, which measures the fraction of states in the lowest Landau level which are occupied by the electrons

\be
\nu=\frac{\Phi_{0}N}{\Phi}\leq 1,\qquad\Phi=\int_\Sigma B\ \ \text{(magnetic flux)},
\ee

where $\Phi_{0}=\frac{2\pi\hslash}{e}$ is the quantum of magnetic flux. In the rest of the paper we choose the natural units $ \hslash=e=1$ so that $\Phi_{0}=2\pi$. These quantum phases of matter encode a new kind of order, the topological order, which is not associated to any local order parameter, but rather to global non-local observables \cite{wen}. Among the observables which capture the topological order of the quantum Hall states, we have the statistics of the topological defects, quasi-holes and quasi-particles,  
which may be inserted at given points $\{w_k\}$ in the surface $\Sigma$ where the electrons move.\\
It has been suggested by C.Vafa a remarkable connection between FQHE and extended supersymmetry which may lead to new insights in the phenomenology of these phases of matter \cite{vafa}. This construction motivates a microscopic description of FQHE in terms of a $\mathcal{N}=4$ supersymmetric Hamiltonian (see \cite{sqm} and \cite{thesis} for a recent review of SQM). Choosing the plane as Riemann surface, we have a Landau-Ginzburg model with superpotential

\begin{equation}\label{vafapoten}
\mathcal{W}(z)= \sum_{i=1}^{N}\left( \sum_{a=1}^{n}\log (z_{i}-x_{a})-\sum_{\ell=1}^{M}\log(z_{i}-\zeta_{\ell}) \right)  + \frac{1}{\nu}\sum_{i<j}\log (z_{i}-z_{j}),
\end{equation}

where $z_{i},i=1,...,N $ are the electron coordinates and $x_{a},\zeta_{\ell}$ are respectively the positions of quasi-holes and magnetic fluxes. The term $\log(z-x_{a})$ is the two dimensional Coulombic potential which describes the interaction between an electron and a magnetic flux unit at $x_{a}$, while the term $\frac{1}{\nu}\sum_{i<j}\log (z_{i}-z_{j})$ keeps track of the Coulomb repulsion between electrons. In order to reproduce the macroscopic magnetic field one should consider a uniform distribution of the flux sources $\zeta_{\ell}$ in $\mathbb{C}$.\\ 
Varying the parameters in $\mathcal{W}$ we get the Berry's connection $D$ on the bundle of vacua which satisfies a set of equations called $tt^{*}$ geometry \cite{rif10}. One can define in terms of $D$ the $tt^{*}$ Lax connection 

\begin{equation}\label{lax}
\mathcal{D}_{\zeta}= D+\frac{1}{\zeta}C, \hspace{1cm} \overline{\mathcal{D}}_{\zeta}= \overline{D}+ \zeta\overline{C},
\end{equation}

where $C,\overline{C}$ denotes the action of chiral and antichiral operators on the vacua and $\zeta \in \mathbb{C}^{\times}$ is an arbitrary parameter. The $tt^{*}$ equations can be rephrased as flatness conditions for $\mathcal{D}_{\zeta},\overline{\mathcal{D}}_{\zeta}$. This connection admits flat sections $\Psi_{\alpha}$ which satisfy \cite{rif23,rif12}

\begin{equation}\label{equations}
\mathcal{D}_{\zeta}\Psi_{\alpha}= \overline{\mathcal{D}}_{\zeta}\Psi_{\alpha}=0.
\end{equation}

According to the Vafa's program, the topological order of FQHE and in particular is captured by the flat connection $\mathcal{D}_{\zeta}$.\\ Some of the main predictions of \cite{vafa} have been verified in \cite{paper}. It has been shown that the degeneracy of the lowest Landau level of an electron coupled to a uniform magnetic field $B$ can be mapped to the degeneracy of vacua of a supersymmetric system with four supercharges. Moreover, it is argued that any Hamiltonian describing the motion in a plane of many electrons coupled to a strong magnetic field is described (at the level of topological order) by the Vafa's $\mathcal{N}=4$ Hamiltonian independently of the details of the other interactions. This result implies that the LG model with superpotential \ref{vafapoten} represents the correct universality class of the many-electron theory. Solving the equations \ref{equations} for this model we find that the $tt^{*}$ connection takes the form of a $SU(2)$ Kohno connection up to a line bundle twisting \cite{paper}

\begin{equation}
\mathcal{D}=d+ \lambda \sum_{i<j}s_{\ell}^{i}s_{\ell}^{j} \frac{d(w_{i}-w_{j})}{w_{i}-w_{j}} + \xi \sum_{i<j}d\log(w_{i}-w_{j})
\end{equation}

where $\xi=-\lambda/4$. This connection acts on the space $V^{n+M}=\bigotimes_{i} V_{i}$ where $V_{i}\simeq \mathbb{C}^{2},i=1,...,n+M$. The operators $s_{\ell}^{i}, \ell=1,2,3$ are the $su(2)$ generators acting on the $V_{i}$ factor, namely 

\begin{equation}
s_{\ell}^{i}= 1 \otimes....\otimes 1 \otimes \frac{1}{2}\sigma_{\ell} \otimes 1 \otimes ....\otimes 1,
\end{equation}

where $\sigma_{\ell}$ are the Pauli matrices. The monodromy representation of the flat connection above is unitary and corresponds to an Hecke algebra representation of the braid group $\mathcal{B}_{n+M}$ which factorizes through the Temperley-Lieb algebra $A_{n+M}(q)$ with \cite{hecke1}

\begin{equation}
q= \exp (\pi i \lambda).
\end{equation}

One can restrict to the monodromy representation of $\mathcal{B}_{n}$ and find the statistics of the quasi-holes.
It turns out that the parameter $\lambda$ is related to the filling fraction $\nu$ by 

\begin{equation}
q^{2}=e^{2\pi i /\nu},
\end{equation}

which gives the two possibilities $q=\pm e^{i\pi/\nu}$. It is believed that $3d$ topological Chern-Simons theories are effective description of FQHE states \cite{wen,tong}. Since the braiding of conformal blocks in $2d$ WZW models is the same of the Wilson lines in non-Abelian Chern-Simons theories, it is natural to require the $tt^{*}$ connection to be a Knizhnik-Zamolodchikov connection for $SU(2)$ current algebra with level $k$ quantized in integral units, namely 

\begin{equation}
\lambda=\pm \frac{2}{\kappa+2}, \hspace{0.5cm} \kappa \in \mathbb{Z}.
\end{equation}

This condition leads to determine the values of the filling fraction which are consistent with $tt^{*}$ geometry. In the case of $q=e^{i\pi/\nu}$ we get

\begin{equation}\label{all1}
\nu= \frac{b}{2 b \pm 1}, \ \ b=\frac{\kappa+2}{2} \geq 1, \ \ \kappa \ \mathrm{even}, \hspace{1.5cm} \nu= \frac{b}{2(b \pm 1)}, \ \ b=\kappa+2\geq 3, \ \ \kappa \ \mathrm{odd},
\end{equation}

where the first one corresponds to the principal series of FQHE. The element $\sigma_{i}^{2}$ of the pure braid group for the principal series has two distinct eigenvalues, in correspondence with the two different fusion channels of the degenerate field $\phi_{1,2}$ in the minimal $(2b, 2b \pm1)$ Virasoro model (see \cite{paper}). This result confirms one of the main prediction of \cite{vafa}, namely the quasi-holes have the same non-Abelian braiding properties of $\phi_{1,2}$ in minimal models. The less natural solution $q=-e^{i\pi/\nu}$ gives other two series of filling fractions. These are respectively

\begin{equation}\label{all2}
\nu= \frac{m}{m+2}, \ \ m=\kappa+2 \geq 2, \hspace{1.5cm} \nu= \frac{m}{3m-2}, \ \ 
m=\kappa+2\geq 2,
\end{equation}

where the first series contains the values of $\nu$ corresponding to the Moore-Read \cite{nab2} and Read-Rezayi models \cite{nab3}. In general there can be overlaps between the series of filling fractions and a certain $\nu$ may appear more than once.\\ In this paper we want to extend the analysis of \cite{paper} and discuss other observables which characterize the topological order of FQHE. The $SU(2)$ Chern-Simons theory has a spectrum of Wilson lines labelled by irreducible representations of $SU(2)$. On the FQHE side, the Wilson lines represent the quasi-holes excitations of a given quantum Hall state, where the filling fraction $\nu$ is related to the Chern-Simons level $\kappa$ according to the above formulas. In the first part of the paper we discuss the $tt^{*}$ solution of the Vafa Hamiltonian for the non-minimal quasi-holes of FQHE. The vacuum bundle of the $N$-particle LG model $\mathcal{V}_{N}$ contains a family of subbundles $\mathcal{V}_{N}^{k_{1},...,k_{d}}$ which are preserved by the parallel transport with the $tt^{*}$ connection. The corresponding fibers $V_{N}^{k_{1},...,k_{d}}$ are subspaces of 

\begin{equation}
V^{k_{1}}\otimes ...... \otimes V^{k_{d}}
\end{equation}

where the factor $V^{k_{i}}$ in the tensor product space denotes the $SU(2)$ irreducible representation with spin $k_{i}/2$. The Hilbert space $V_{N}^{k_{1},...,k_{d}}$ corresponds to the anyonic fusion space for the set of quasi-holes with spins $k_{1}/2,...,k_{d}/2$, where the minimal quasi-hole is associated to the spin $1/2$ representation. Each subbundle $\mathcal{V}_{N}^{k_{1},...,k_{d}}$ is endowed with the $SU(2)$ Knizhnik-Zamolodchikov connection specialized to the corresponding representation.\\ In the second part of the paper we discuss the FQHE on a generic Riemann surface. An interesting topological invariant which captures the quantum order of FQHE (and topological phases in general) is the degeneracy of the ground state. This must depend only on the filling fraction $\nu$ labelling the quantum Hall states and the genus $g$ of the Riemann surface on which the electrons are confined. By consistency with the $3d$ topological field theory, the ground state degeneracy should be the same of the $SU(2)$ Chern-Simons theory on a $g$-Riemann surface at the level $\kappa$ corresponding to $\nu$. This is computed by the well known Verlinde formula \cite{verlinde}

\begin{equation}
\mathrm{deg}_{g,\kappa}= \left( \frac{\kappa+2}{2}\right) ^{g-1} \sum_{n=0}^{\kappa} \left( \sin \frac{(n+1)\pi}{\kappa+2}\right)^{2(1-g)} , \ \ \  g\geq 1.
\end{equation}

It is straightforward to generalize the construction of the Vafa Hamiltonian on higher genus Riemann surfaces and guess the corresponding $tt^{*}$ solution. This should correspond to the higher genus generalization of the Knizhnik-Zamolodchikov connection, also called Knizhnik-Zamolodchikov-Bernard connection \cite{bern1,bern2,kzb}, which is known to provide the monodromy representation of conformal blocks in $2d$ WZW models. 

\section{Non-Minimal Quasi-Holes}\label{nonminimalsec}

\subsection{The Landau-Ginzburg Model}

We consider $N$ electrons moving on the plane $\mathbb{C}$. We denote with $z_{i}, i=1,...,N$ and $x_{a}, a=1,...,n$ the positions of electrons and quasi-holes respectively, and with $\zeta_{\ell},\ell=1,...,M$ the positions of the magnetic fluxes. We take the $d=M+n$ points $\zeta_{\ell},x_{a}$ all distinct. The superpotential of the LG model is 

\begin{equation}\label{formby}
\begin{split}
&\mathcal{W}(z_{i})=   \sum_{i=1}^{N}W(z_{i})+ \beta\sum_{i<j}\log (z_{i}-z_{j})^{2}  ,\\ 
& W(z)= \mu z + W_{\mathrm{Vafa}}(z), \hspace{1cm} W_{\mathrm{Vafa}}(z)=\sum_{a=1}^{n}k_{a}\log (z-x_{a}) - \sum_{\ell=1}^{M}k_{\ell} \log(z-\zeta_{\ell}).
\end{split} 
\end{equation}

As explained in \cite{paper}, it is convenient to introduce the coupling $\mu$ to make the problem better behaved. In particular, this modification of the Vafa superpotential ensures the normalizability of the vacuum wave functions and the existence of an energy gap between the ground state and the first excited level of the Hamiltonian. From the point of view $2d$ electrostatic, the coupling $\mu$ plays the role of background electrostatic field. As long as it is non zero, it can be fixed to the value that we prefer, since the $tt^{*}$ monodromy is indipendent from the parameters. The residues of $\mathcal{W}$ at its poles are fixed by physical considerations. The real part of $W_{\mathrm{Vafa}}(z)$ has the interpretation of electrostatic potential of a system of point-like charges. The punctures in $\zeta_{\ell}$ form a regular distribution to reproduce the macroscopic magnetic field and carry $k_{\ell} \in \mathbb{N}$ units of magnetic flux. The mignus sign in front of the logarithmic interaction is related to the choice of the orientation. The quasi-holes in $x_{a}$ are just magnetic fluxes with opposite sign and size $k_{a} \in \mathbb{N}$. The superpotential \ref{vafapoten} describes minimal quasi-holes and magnetic fluxes which carry $\pm 1$ flux respectively. The remaining residue is the the size of the Coulomb repulsion $\beta$. Working in a periodic box, this coupling is frozen to the rational number $1/2\nu$ (see \cite{paper} and section \ref{riemann}). Despite the above considerations, since the monodromy representation is independent of the couplings and the model is well defined for any value of the residues, we are free to deform them away from their physical values according to convenience. The non-frozen couplings are the $x_{a}$ and the $\zeta_{\ell}$. These form a set of $d$ distinct points in $\mathbb{C}$ in which are identified modulo permutations the ones with equal charge. In the most general case, the manifold of essential couplings is the space of $d$ ordered points

\begin{equation}
\mathcal{C}_{d}=\left\lbrace w_{1},....,w_{d} \in \mathbb{C}^{d} \ \vert \ w_{i}\neq w_{j} \ \mathrm{for} \ i\neq j \right\rbrace,
\end{equation}

where $w_{i}=x_{a},\zeta_{\ell}$. The fundamental group $P_{d}=\pi_{1}(C_{d})$ is called the pure braid group of $d$ strings.\\ 
The fundamental degrees of freedom of the Vafa model are not the fields $z_{i}$, but the symmetric polynomials

\begin{equation}
e_{k}= \sum_{1\leq i_{1}<....<i_{k}\leq N} z_{i_{1}}....z_{i_{k}}. 
\end{equation}

Indeed the superpotential is manifestly invariant under $S_{N}$ permutations of the $z_{i}$ and can be rewritten in these coordinates. The target manifold of the theory is the configuration space of $N$ identical particles on $\mathbb{C}\setminus\left\lbrace x_{a},\zeta_{\ell} \right\rbrace $, i.e. 

\begin{equation}\label{space}
X_{d,N}=\left\lbrace (z_{1},...,z_{N}) \in \left( \mathbb{C}\setminus \lbrace x_{a},\zeta_{\ell}\rbrace \right)^{N} \big\vert z_{i}\neq z_{j}, \  \mathrm{for} \ i\neq j \right\rbrace \big/ S_{N}.
\end{equation} 

The classical vacua of the theory are the solutions to the equations $\partial_{e_{k}}\mathcal{W}(e_{j})=0$. It is known that for generic values of the couplings the zeores of $d\mathcal{W}$ are isolated and non-degenerate with multiplicity \cite{rif23,paper}

\begin{equation}\label{bose}
r_{d,N}= \begin{pmatrix}
N+d-1 \\ N
\end{pmatrix}.
\end{equation}

The Hilbert space $\mathbf{H}$ of the model is the space of differential forms $\psi$ on the target manifold $X_{d,N}$ with $L^2$-coefficients \cite{sqm}. The Lagrangian of the theory is invariant under a supercharge $\overline{Q}$ which acts
on forms as 

\be\label{Qbar}
\overline{Q}\psi=\overline{\partial}\psi+d\mathcal{W}\wedge\psi.\ee

The operator $\overline{Q}$ is nilpotent, i.e. $\overline{Q}^2=0$, and commutes with multiplication by holomorphic functions. The vacuum vector space
\be\label{fiber}
V:=\Big\{\psi\in \mathbf{H}\; :\; \overline{Q}\psi=\overline{Q}^\dagger \psi=0\Big\}\subset\mathbf{H}
\ee is isomorphic to the cohomology of $\overline{Q}$ in $\mathbf{H}$. Since the classical vacua are isolated, the vacuum space $V$ consists of primitive forms of degree $N=\mathrm{dim} \ X_{d,N}$ (see theorem in \cite{sqm}). Moreover, the dimension of the ground state of the quantum system is equal to the number of classical vacua counted with multiplicity, namely $r_{d,N}=\mathrm{dim} \ V$, and coincides with the Witten index of the SQM model. This number is invariant under continuous deformations of the couplings such that $\vert d\mathcal{W}\vert ^2$ remains bounded away from zero outside a large compact set $C \subset  X_{d,N} $.

\subsection{Generalites of the $tt^{*}$ Monodromy Representation}

Let $\zeta\in\bP^1$ and consider the smooth function
\be
F(z_{i},\overline{z}_{i};\zeta)=\mathrm{Re}\big(\mathcal{W}(z_{i})/\zeta+\overline{\mathcal{W}}(\overline{z}_{i})\zeta\big).
\ee
Morse cobordism implies the isomorphism \cite{morse}
\be\label{lllasqwp}
V\cong H^*(X_{d,N}, X_{d,N}^{\Lambda,\zeta};\C),
\ee
where $H^*(X_{d,N}, X_{d,N}^{\Lambda,\zeta};\C)$ denotes the relative cohomology with complex coefficients, and 
\be
X_{d,N}^{\Lambda, \zeta}:=\Big\{z_{i}\in X_{d,N}\;:\; F(z_{i},\bar z_{i};\zeta)> \Lambda\Big\}\subset X_{d,N}
\ee
for some sufficiently large constant $\Lambda$. Since the vacuum space is spanned by $N$-forms, the space $H^\ast(X_{d,N}, X_{d,N}^{\Lambda,\zeta};\Z)$ is non-zero only in degree $N$. The dual relative homology $H_*(X_{d,N}, X_{d,N}^{\Lambda,\zeta};\C)$ is called the space of branes because in $2d$ the corresponding objects have the physical interpretation of half-BPS branes \cite{morse}; the twistor parameter $\zeta$ specifies which linear combinations of the original 4 supercharges are preserved by the brane. The space of branes has an integral structure given by homology with integral coefficients
\be\label{nnnncx}
V\cong H_\ast(X_{d,N}, X_{d,N}^{\Lambda,\zeta};\Z)\otimes_\Z \C.
\ee
An integral basis of $H_\ast(X_{d,N}, X_{d,N}^{\Lambda,\zeta};\Z)$ is given by special Lagrangian submanifolds of $X_{d,N}$ and, more specifically, by Lefshetz timbles describing the gradient flow of $F(z_{i},\overline{z}_{i};\zeta)$ for generic $\zeta$ \cite{morse}. Denoting with $B_{\alpha,\zeta}$, ($\alpha=1,\dots,r_{d,N}$) such an integral basis and with $\psi_{j}$
($j=1,\dots, r_{d,N}$) a basis of $V$, we may form the non-degenerate
$r_{d,N}\times r_{d,N}$ matrix
\be\label{brane}
\Psi(\zeta)_{j\alpha}=\langle\psi_j| B_{\alpha,\zeta} \rangle= \int_{B_{\alpha,\zeta}} e^{-\mathcal{W}/\zeta+\overline{\mathcal{W}}\zeta } \, \psi_j
\ee
called the brane amplitudes.\\ Over the space of couplings $C_{d}$ we can construct the vacuum vector bundle

\begin{equation}
\mathcal{V}\rightarrow C_{d}
\end{equation}

namely the subbundle of the trivial Hilbert bundle $\mathbf{H} \times C_{d}$ whose fiber is the vacuum space $V$. The vacuum Berry connection, induced on the bundle of vacua by the trivial connection of the Hilbert bundle, is metric with respect to the $tt^{*}$ hermitian scalar product

\begin{equation}
g_{i\bar{j}}= \int_{X_{d,N}} \psi_{i} \wedge \ast \overline{\psi_{j}}
\end{equation}

and preserves the holomorphic structure of the vacuum bundle. There is a unique such connection, namely the Chern connection of $g$, whose $(1,0)$ and $(0,1)$ parts in an holomorphic frame of $\mathcal{V}$ are \cite{rif10}

\begin{equation}
D=\partial + g\partial g^{-1}, \hspace{1cm} \overline{D}=\overline{\partial}.
\end{equation}

We can further define a family of flat connections parametrized by $\zeta$, the $tt^{*}$ Lax connection \cite{rif12}

\begin{equation}
\begin{split}
& \mathcal{D}_{\zeta}= D+\zeta^{-1} C, \\ \\ & \overline{\mathcal{D}}_{\zeta}= \bar{\partial}+\zeta \overline{C},
\end{split}
\end{equation}

where $C$ is a $(1,0)$ form describing the action of the chiral operators on the vacua and $\overline{C}$ is the hermitian conjugate with repsect to the metric $g$. The $tt^{*}$ equations prescribe the curvature of the Berry connection and can be rephrased as the flatness conditions for the Lax connection, i.e. \cite{rif10,rif12}

\begin{equation}
(\mathcal{D}_{\zeta})^{2}= (\overline{\mathcal{D}}_{\zeta})^{2} =  \mathcal{D}_{\zeta} \overline{\mathcal{D}}_{\zeta} + \overline{\mathcal{D}}_{\zeta} \mathcal{D}_{\zeta}=0.
\end{equation}

The columns of the $r_{d,N} \times r_{d,N}$ matrix \ref{brane} form a basis of sections of the vacuum bundle satisfying the system of linear equations

\begin{equation}\label{linprobl}
\mathcal{D}_{\zeta}\Psi_{\alpha}(\zeta)=\overline{\mathcal{D}}_{\zeta}\Psi_{\alpha}(\zeta)=0,
\end{equation}

also known as Lax equations. Taking the analytic continuation of the solutions along a non trivial loop $\gamma$ in the space of couplings $C_{d}$ we obtain the monodromy $\rho(\gamma)$ defined by

\begin{equation}
\Psi(w_{i},\zeta)\rightarrow \Psi(w_{i},\zeta)\rho(\gamma).
\end{equation}

Since the branes are representatives of integral homology classes, for each $\zeta \in \mathbb{P}^{1}$ they define a local system on $C_{d}$ canonically equipped with a flat connection, the Gauss-Manin one. Dually, the brane amplitudes define the $\mathbb{P}^{1}$-family of $tt^{*}$ Lax connections. The natural identification of the two connections implies that the $tt^{*}$ monodromy representation $\rho$ may be conjugated such that 

\begin{equation}
\rho: \pi_{1}(C_{d})\rightarrow SL(r_{d,N},\mathbb{Z}).
\end{equation}

Since the entries of the matrix $\rho(\gamma)$ and its inverse are integers, they are locally indipendent from the couplings and the spectral parameter $\zeta$.\\ 
The main problem is to compute the monodromy representation of the $tt^{*}$ Lax connection. This provides the boundary condition to solve the $tt^{*}$ equations. Following \cite{paper}, it is natural to consider the limit in which we rescale the superpotential 

\begin{equation}
\mathcal{W}\rightarrow R \mathcal{W}
\end{equation}

and send $R\rightarrow 0$ while the spectral parameter $\zeta$ remains fixed. In the two dimensional $\mathcal{N}=(2,2)$ version of the model the parameter $R$ plays the role of RG scale and the limit $R\rightarrow 0$ corresponds to the UV limit of the theory. In this limit the Lax connection reduces to the Berry connection

\begin{equation}
\mathcal{D}_{\zeta} +\overline{\mathcal{D}}_{\zeta} \xrightarrow{R \rightarrow 0, \ \zeta \ \mathrm{fixed} } D + \overline{D}.
\end{equation}

In particular, the Lax connection is indipendent from the spectral parameter $\zeta$ in the UV limit. The Lax connection is the one that captures the topological order of FQHE, while the Berry connection is the one that provides the unitary time evolution of the quantum system. Hence, the UV limit in which the two coincide turns out to be the suitable one to FQHE. 

\subsection{$\theta$-vacua and filling fraction}

We note that the superpotential of the Vafa model is a multivalued function. This is not an issue, since the Hamiltonian contains only the first and second derivatives of $\mathcal{W}$, which are well defined meromorphic functions on $X_{d,N}$. The $1$-form $d\mathcal{W}$ is closed but not exact, since the target manifold of the theory is not simply connected. The fundamental group is 

\begin{equation}
\pi_{1}(X_{d,N})= \mathcal{B}(N,S_{0,d+1}),
\end{equation}

where $\mathcal{B}(N,S_{g,d+1}) $ is the braid group of $N$ strings on the surface $S_{g,d+1}$ of genus $g$ with $d+1$ punctures. The group $\mathcal{B}(N,S_{0,d+1})$ has the convenient presentation \cite{belling}

\begin{equation}
\begin{split}
& \mathrm{generators}:  \sigma_{1},\sigma_{2},...,\sigma_{N-1},z_{1},z_{2},...,z_{d} \\ 
& \mathrm{relations}: \begin{cases} \sigma_{i}\sigma_{i+1}\sigma_{i}= \sigma_{i+1}\sigma_{i}\sigma_{i+1}, \hspace{2cm} \sigma_{i}\sigma_{j}=\sigma_{j}\sigma_{i} \ \ \mathrm{for} \vert i-j \vert \geq 2, \\ 
z_{j}\sigma_{i}=\sigma_{i}z_{j} \ \mathrm{for}\ i\neq 1, \hspace{2.3 cm} \sigma_{1}^{-1}z_{j}\sigma_{1}^{-1}z_{j}= z_{j}\sigma_{1}^{-1}z_{j}\sigma_{1}^{-1}, \\ 
\sigma_{1}^{-1}z_{j}\sigma_{1}^{-1}z_{l}= z_{l}\sigma_{1}^{-1}z_{j}\sigma_{1}^{-1} \ \ \mathrm{for} \ j<l.
\end{cases}
\end{split}
\end{equation}

The $\sigma_{i}$ generate a subgroup of $\mathcal{B}(N,S_{0,d+1})$ which is isomorphic to the Artin Braid group $\mathcal{B}_{N}$. The abelianized Galois group $\mathcal{B}(N,S_{0,d+1})^{\mathrm{Ab}}\simeq \mathbb{Z}^{d+1}$ is the homology group of the target space generated by $\sigma,z_{1},...,z_{d}$. The dual cohomology group $H^{1}(X_{d,N},\mathbb{Z})$ has the generators 

\begin{equation}
\frac{1}{2\pi i} d \log \prod_{i<j}(z_{i}-z_{j})^{2}, \hspace{1cm} \frac{1}{2\pi i }d\log \prod_{i,a}(z_{i}-x_{a}), \hspace{1cm} \frac{1}{2\pi i} d\log \prod_{i,\ell}(z_{i}-\zeta_{\ell}).
\end{equation}

To keep track of the non trivial topology of the target space we may pull-back the Landau-Ginzburg model on the abelian universal cover $\mathcal{A}_{d,N}$ \cite{paper}. Being a symmetry of the model on $\mathcal{A}_{d,N}$, the vacuum space decomposes in a direct sum of unitary representations of the homology group

\begin{equation}
V_{\mathcal{A}}= \bigoplus_{\chi \in \mathrm{Hom}\left( H^{1}(X_{d,N},\mathbb{Z}),U(1)\right) } V_{\chi}, \hspace{1cm} \mathrm{dim} \ V_{\chi}=r_{d,N}.
\end{equation}

Identifying $H_{1}(X_{d,N};\mathbb{Z})$ with $\mathbb{Z}^{d+1}$, the representations $V_{\chi}$ are in correspondence with the characters

\begin{equation}
\chi_{\theta, \phi_{a}, \varphi_{\alpha}}: \vec{n}\rightarrow e^{i n\theta + i\sum_{a} n_{a}\phi_{a} + i \sum_{\ell}n_{\ell}\varphi_{\ell}}
\end{equation}

labelled by the set of angles $\theta, \phi_{a}, \varphi_{\ell} \in [0,2\pi]$. The model on $\mathcal{A}_{d,N}$ has its own BPS branes which pair with the vacua of the ground state and form a basis of sections for the vacuum bundle $\mathcal{V}_{\mathcal{A}}$ of the covering model. Let $q=e^{i\theta}, q_{a}=e^{i \phi_{a}}, q_{\ell}=e^{i \varphi_{\ell}}$. The isomorphism

\begin{equation}
V_{\mathcal{A}}\cong H_{*}(\mathcal{A}_{d,N},\mathcal{A}^{\Lambda,\zeta}_{d,N},\mathbb{Z}) \otimes_{\mathbb{Z}} \mathbb{C}
\end{equation}

provides the ground state with the structure of a free $\mathbb{Z}\left[ \left\lbrace q^{\pm 1}\right\rbrace ,\left\lbrace q_{a}^{\pm 1}\right\rbrace , \left\lbrace q_{\ell}^{\pm 1}\right\rbrace \right] $-module of rank $r_{d,N}$ \cite{paper}. \\ In the physical FQHE the punctures $\zeta_{\ell}$ reproduce the macroscopic magnetic field $B$ and carry an integral number of magnetic flux units. On the other hand, the quasi-holes are just magnetic fluxes with opposite charge. Hence, interpreting the angles $\phi_{a}, \varphi_{\ell}$ as Aharonov-Bohm phases, we set 

\begin{equation}
\phi_{a}= 0 \ \mathrm{mod} \ 2\pi , \hspace{1cm} \varphi_{\ell}= 0 \ \mathrm{mod} \ 2\pi.
\end{equation}

It follows that the model describing the physical FQHE should be defined on the abelian subcover corresponding to the unique non trivial angle $\theta$. Following the discussion in \cite{paper}, in order to have normalizable states we should ask $\theta$ to be a rational multiple of $2\pi$

\begin{equation}
\theta= \pi \left( 1+\frac{a}{b}\right) , \hspace{1cm} a \in \mathbb{Z}, b \in \mathbb{N}, \hspace{0.5cm} \big\vert \ \mathrm{gcd}(a,b)=1 \hspace{0.5cm} \mathrm{and} \ -b\leq a \leq b.
\end{equation}

Switching on a non-zero $\theta$ corresponds to insert in the BPS brane amplitudes \ref{brane} the chiral field \cite{paper}

\be
\prod_{i<j}(z_i-z_j)^{\theta/\pi}.
\ee

Keeping into account the Jacobian arising from the change of variable $\{z_i\}\mapsto\{e_k\}$, the vacuum wave-functions in terms of the $z_i$'s contain the factor

\be
\prod_{i<j}(z_i-z_j)^{1+\theta/\pi},\qquad 0\leq \theta \leq 2\pi.
\ee  

The comparison with the Laughlin wave-functions \cite{wen,tong} leads to the identification

\be\label{nutheta}
\frac{1}{\nu}=1+\frac{\theta}{\pi}=2\pm \frac{a}{b}
\ee

which gives $1\leq 1/\nu\leq 3$. The interpretation of $1/\nu$ as $\theta$-angle appears naturally in the context of the supersymmetric description of FQHE and reflects the connection between topological order and the non-local entanglement of the electrons. The UV Berry connection perserves the $\theta$-sectors of the vacuum space and defines a group homomorphism 

\begin{equation}
\rho_{\theta}: \pi_{1}(C_{d})\rightarrow GL (r_{d,N}, \mathbb{Z}\left[ \left\lbrace q^{\pm 1}\right\rbrace \right] ).
\end{equation}

In other words, the entries of the monodromy matrices are Laurent polynomials in the variable $q$ with integer coefficients.

\subsection{Subbundle Decomposition}

We observe that the multiplicity of vacua $r_{d,N}$ corresponds to the Bose statistics. The interpretations of this result is clear in the limit of $\beta\rightarrow 0$: in a classical vacuum configuration of $\mathcal{W}$ the $z_{i}$ are close to the vacua of the single particle model $W(z)$ and several of them may take different values around the same one-field vacuum. Since these values differ by orders $O(\beta)$, in the $\beta\rightarrow 0$ limit the vacuum configurations are simply labelled by a set of positive integers $(n_{1},....,n_{d})$ which denote how many particles we put in the vacua of the single field model. \\ 
It turns out that the vacuum bundle of the $N$-particle LG model decomposes in subbundles which are preserved by parallel transport with the $tt^{*}$ flat connection. This decomposition has been studied by Gaiotto and Witten in \cite{GW} from the homological point of view. Following their argument, since the monodromy representation is independent of $\mu$ as long as it is non-zero, we can take it very large while keeping $\beta$ finite. This is also the resonable regime of FQHE. Up to $ O(1/\mu)$ corrections, in a vacuum configuration $\lbrace z_{i} \rbrace_{i=1,...,N} $ which solves the equations

\begin{equation}
\sum_{\ell} \frac{k_{\ell}}{z_{i}-\zeta_{\ell}} - \sum_{a} \frac{k_{a}}{z_{i}-x_{a}} = \mu + \sum_{j \neq i } \frac{2\beta}{z_{i}-z_{j}}
\end{equation}

the $z_{i}$ are approximately equal to one of the punctures $x_{a},\zeta_{\ell}$ and the product of $N$ one particle Lefschetz thimbles is approximatively a brane for the full interacting model. Despite the actual brane differs from the product of one-particle ones by some $O(1/\mu)$ correction, they belong to the same homology class and this is enough for monodromy considerations. This provides a recipe to construct a basis for the space of branes of the LG model. 
Let 

\begin{equation}
\left\lbrace w_{1},...,w_{d} \right\rbrace = \left\lbrace x_{1},....,x_{n}, \zeta_{1},...,\zeta_{M} \right\rbrace,
\end{equation}

where $d=M+n$, and denote with $B_{i_{1},...,i_{N}}=B_{i_{1}}\times.....\times B_{i_{N}}$ a brane of the $N$ particle model as product of single particle branes. In the limit of large $\mu$, the cycle $B_{i}$ associated to the vacuum configuration $w_{i}+ O(1/\mu)$ is approximately a straight line on the complex plane starting at $w_{i}$. We consider the $N$-particle Lefschetz thimbles in which there are at most $k_{i} \in \mathbb{N}$ one particle cycles starting at $w_{i}$. The integers $k_{i}$ must satisfy $1 \leq k_{i} \leq N$ and $N\leq\sum_{i} k_{i}$. The subspaces $B_{N}^{k_{1},...,k_{d}}$ of the relative homology spanned by these branes are left invariant by arbitrary braidings of the punctures $w_{i}$ \cite{GW}. Given the identification between the $tt^{*}$ flat connection and the Gauss-Manin connection of the local system of BPS branes, one concludes that the vacuum bundle of the LG model contains a collection of invariant subbundles $\mathcal{V}_{N}^{k_{1},...,k_{d}}$ whose fibers $V_{N}^{k_{1},...,k_{d}} $, dual to $B_{N}^{k_{1},...,k_{d}}$, define a family of subrepresentations of the $tt^{*}$ monodromy. 
The rank of the subbundle $\mathcal{V}_{N}^{k_{1},...,k_{d}}$ is defined by the $(k_{1},...,k_{d})$-statistics

\begin{equation}
\mathrm{rank} \ \mathcal{V}_{N}^{k_{1},...,k_{d}}= \# \,(n_{1},...,n_{d}) \  \bigg \vert  \ 0\leqslant n_{i} \leqslant k_{i}\  \mathrm{and} \ \sum_{i=1}^{d}n_{i}=N,
\end{equation}

where in the vacuum configuration $(n_{1},...,n_{d})$ we dispose $n_{i}$ electrons in the single particle vacuum $z_{i}=w_{i}+ O(1/\mu)$. The minimal quasi-holes and fluxes cannot host more than one particle and the dimension of the corresponding Hilbert space reduces to the Fermi statistics

\begin{equation}
\mathrm{rank} \ \mathcal{V}_{N}^{1,...,1}=  \begin{pmatrix}
d \\ N
\end{pmatrix}.
\end{equation}

On the other hand, the subspace $\mathcal{V}_{N}^{N,...,N}$ corresponds to the full Hilbert space and the dimension is equal to the Bose counting $r_{d,N}$. The maximum number of particles that we can place in the single particle vacua with $(k_{1},...,k_{d})$-statistics is

\begin{equation}
K=\sum_{i=1}^{d} k_{i}.
\end{equation}

Since a small number $n\ll M$ of topological defects carry a magnetic flux with opposite sign, we have a small mismatch between $K$ and the effective magnetic flux measured by the fall off of the wave function at infinity. In the physical FQHE the number of electrons should satisfy $N=\nu K$ for a given quantum Hall state with filling fraction $\nu\leq 1 $. However, since our considerations apply to any number of particles, we can keep $N$ generic in our analysis.

\subsection{Conformal Blocks Bundle}\label{spindeg}

In a given FQHE topological phase, from the dynamics of the microscopic degrees of freedom there emerges at low-energy an effective 2d QFT $\mathcal{Q}$ for the non-local quasi-hole operators $\Sigma(w)$. The topological order of the phase is captured by the braiding properties of their multi-point correlators

\begin{equation}
\langle \Sigma(w_{1})\Sigma(w_{2})\cdot\cdot\cdot \Sigma(w_{d})\rangle_{\mathcal{Q}}
\end{equation}

as we transport the $\Sigma(w)$'s around each other in closed loops. In connecting the $tt^{*}$ monodromy of the Vafa model with the braid representations of $\mathcal{Q}$-correlators, the brane amplitudes are expressed as ratios of $d$-point functions \cite{paper}. In the UV limit of the $\mathcal{Q}$ theory, which coincides with the physical UV limit of the $2d$ LG model, the multivalued correlation functions become sums of products of left/right conformal blocks. It follows that the $tt^{*}$ flat sections become in the UV limit some combinations of conformal blocks and the equations that they satisfy, the Lax equations, are related in a simple way to the isomonodromic PDEs for the $\mathcal{Q}$-blocks. Both sets of equations define flat connections and monodromy representations. The precise dictionary between the two is provided in section \ref{dictionary}.\\ The geometry of conformal blocks for the minimal quasi-hole operators has been studied in \cite{paper}. For a given puncture at position $w_{j} \in \mathbb{C}$ one introduces the two-component operator

\begin{equation}
\Sigma_{j,\alpha}(w_{j})= \begin{pmatrix}
\sigma_{j}(w_{j}) \\ \mu_{j}(w_{j})
\end{pmatrix}.
\end{equation}

A local basis of sections for the subbundle $\mathcal{V}_{N}^{1,...,1}$ is given by the $\mathcal{Q}$-amplitudes

\begin{equation}\label{amplitudes}
\langle \Sigma_{1,\alpha}(w_{1})\cdot\cdot\cdot \Sigma_{d,\alpha}(w_{d})\rangle  \in V^{\otimes d}
\end{equation}

where $V\simeq \mathbb{C}^{2}$ is the fundamental representation of $SU(2)$. The operators $\mu_{j},\sigma_{j}$ have the same local behaviour of the Ising order and disorder operators, but globally they have in general different braiding properties. The minimal quasi-hole can host at most one particle and is associated with a spin $1/2$ degree of freedom. If a spin up/down operator $\sigma_{j}/\mu_{j}$ is inserted in the position $w_{j}$, the corresponding state is filled/empty. This construction may be generalized to the case of non-minimal quasi-holes. The spectrum of the $\mathcal{Q}$-theory should contain generalized quasi-hole operators 

\begin{equation}
\Sigma^{k_{j}}_{j,m_{j}}(w_{j})= \begin{pmatrix}
\sigma^{k_{j}}_{j,k_{j}/2}(w_{j}) \\ \sigma^{k_{j}}_{j,(k_{j}-1)/2}(w_{j}) \\ \cdot \\ \cdot \\ \cdot \\ \sigma^{k_{j}}_{j,-k_{j}/2}(w_{j})
\end{pmatrix}, \hspace{1cm} -k_{j}/2 \leq m_{j} \leq k_{j}/2, \ \ \ m_{j} \in \frac{1}{2}\mathbb{Z}.
\end{equation}

Analogously to the minimal case, a basis of sections for the subbundle $\mathcal{V}_{N}^{k_{1},..., k_{d}}$ is provided by the $\mathcal{Q}$-conformal blocks

\begin{equation}\label{qampl}
\langle \Sigma^{k_{1}}_{1,m_{1}}(w_{1})\cdot\cdot\cdot \Sigma^{k_{d}}_{d,m_{d}}(w_{d})\rangle  \in V^{k_{1}} \otimes \cdot\cdot\cdot\otimes V^{k_{d}}
\end{equation}

where $V^{k_{j}} \simeq \mathbb{C}^{k_{j}+1}$, $j=1,...,d$ denotes the spin $k_{j}/2$ representation of $SU(2)$. In other words, a non-minimal quasi-hole with charge $k_{j}$ carries a spin $k_{j}/2$ degree of freedom and the insertion in the correlator of the operator $\sigma^{k_{j}}_{j,m_{j}}$ at position $w_{j}$ corresponds to place $m_{j}+k_{j}/2$ electrons in the corresponding single particle vacuum.\\  Proceeding as in \cite{paper}, one may associate to a set of $d$ topological defects with charges $k_{1},...,k_{d}$ the Gran bundle

\begin{equation}\label{sum}
\mathcal{V}^{k_{1},..., k_{d}}=\bigoplus_{N=0}^{K} \mathcal{V}_{N}^{k_{1},..., k_{d}}, \hspace{2cm} K=\sum_{i=1}^{d}k_{i},
\end{equation}

whose fiber is spanned by the $\mathcal{Q}$-amplitudes \ref{qampl} and is modelled on the tensor product space

\begin{equation}
V^{k_{1},..., k_{d}}= V^{k_{1}} \otimes \cdot\cdot\cdot \otimes V^{k_{d}}, \hspace{1cm} \mathrm{dim} \ V^{k_{1},..., k_{d}}= \prod_{j=1}^{d} (k_{j}+1).
\end{equation}

Let $s_{\ell}^{j}$, $\ell=1,2,3,$ be the generators of $su(2)$ in the $k_{j}/2$ representation and

\begin{equation}
L_{\ell}= \sum_{j=1}^{d} 1\otimes \cdot\cdot\cdot \otimes 1 \otimes s_{\ell}^{j} \otimes 1 \otimes \cdot\cdot\cdot \otimes 1
\end{equation}

the $\ell$-component of the total angular momentum. The occupation number $\hat{N}_{j}$ of the $j$-th vacuum and the total particle number $\hat{N}$ are related to the $su(2)$ generators by 

\begin{equation}
\hat{N}_{j}= s_{3}^{j}+k_{j}/2 ,\hspace{1cm} \hat{N}= L_{3} + K/2.
\end{equation}

The $tt^{*}$ geometry of the Gran bundles $\mathcal{V}^{k_{1},..., k_{d}}$ is fully determined by the action of the pure braid group $P_{d}$ on the quasi-hole operators $\Sigma^{k_{j}}_{j,m_{j}}(w_{j})$. The conformal blocks \ref{qampl} are holomorphic multivalued functions on $C_{d}$ which undergo analytic continuation as we move the position of a quasi-hole insertion around another one. The typical monodromy action on conformal blocks of a $2d$ CFT is a specialization of the universal Kohno monodromy (see \cite{hecke1,hecke2,hecke3,BPZ}). For each $d\geqslant 2$ we introduce the Kohno-Drinfield Lie algebra $\mathbf{t}_{d}$ as the Lie algebra generated by the operators $B_{ij}=B_{ji}$, $i,j=1,...,d$, which satisfy the relations \cite{drinkon}

\begin{equation}\label{purebraidrel}
\begin{split}
& \left[ B_{ik},B_{ij} + B_{jk}\right] =0, \hspace{1cm} (i,j,k \ \mathrm{distinct}), \\ 
& \left[ B_{ij}, B_{k\ell} \right]=0, \hspace{1cm} (i,j,k,\ell \ \mathrm{distinct}),
\end{split}
\end{equation}

also known as infinitesimal pure braid relations. A Kohno connection is a meromorphic flat connection $\mathcal{D}=d+\mathcal{A}$ on the configuration space $C_{d}$ with values in $\mathbf{t}_{d}$. It is given by the formula

\begin{equation}
\mathcal{D}=d + \sum_{i<j} \frac{B_{ij}}{w_{i}-w_{j}}d(w_{i}-w_{j}).
\end{equation}

One can show that a connection of this form is flat \cite{drinkon}, i.e. $\mathcal{D}^{2}=0$, if and only if the operators $B_{ij}$ satisfy the relations \ref{purebraidrel}.\\ The UV Berry connection acting on the $\mathcal{Q}$-conformal blocks should take the form a Kohno connection. By an homomorphism of Lie algebras 

\begin{equation}\label{map}
\Upsilon_{d}: \mathbf{t}_{d} \rightarrow \mathrm{End}\left( V^{k_{1}}\otimes \cdot\cdot\cdot \otimes V^{k_{d}} \right)  
\end{equation}

the braid generators $B_{ij}$ are represented by constant matrices acting on $V^{k_{1}}\otimes \cdot\cdot\cdot \otimes V^{k_{d}}$. Each Gran bundle $\mathcal{V}^{k_{1},..., k_{d}}$ is therefore equipped with the UV Berry connection specialized to the corresponding representation. By parallel transport with $\mathcal{D}$, the vector spaces $V^{k_{1},..., k_{d}}$ gain the structure of monodromy representations of the pure braid group $P_{d}$. The matrices $B_{ij}$ have been computed in \cite{paper} for the case in which all the punctures are in the fundamental representation of $su(2)$. The generalization of $\mathcal{D}$ to arbitrary $(k_{1},...,k_{d})$-representations is straightforward. We introduce the $su(2)^{\otimes 2}$ operator $\Omega= \sum_{\ell=1}^{3} s_{\ell}\otimes s_{\ell} $ and denote with

\begin{equation}
\Omega_{ij}= \sum_{\ell=1}^{3} 1\otimes \cdot\cdot\cdot \otimes 1 \otimes s_{\ell}^{i} \otimes 1 \otimes \cdot\cdot\cdot \otimes 1 \otimes s_{\ell}^{j} \otimes 1 \otimes\cdot\cdot\cdot \otimes 1
\end{equation}

the action of $\Omega$ on the $i$-th and $j$-th factor of $V^{k_{1}}\otimes \cdot\cdot\cdot \otimes V^{k_{d}}$. The explicit form of the map \ref{map} is given by

\begin{equation}
\Upsilon_{d}: B_{ij}\rightarrow \lambda(\theta) \, \Omega_{ij}.
\end{equation}

The general solution of the $tt^{*}$ equations is therefore the $su(2)$ Knizhnik-Zamolodchikov connection 

\begin{equation}
\mathcal{D}= d + \lambda(\theta)\sum_{i<j} \frac{\Omega_{ij}}{w_{i}-w_{j}} d(w_{i}-w_{j}).
\end{equation}

One can check that the operators $\Omega_{ij} \in \mathrm{End}(V^{k_{1}}\otimes \cdot\cdot\cdot \otimes V^{k_{d}})$ satisfy the infinitesimal pure braid relations \ref{purebraidrel}, ensuring the integrability of the connection \cite{drinkon}. The parameter $\lambda(\theta)$ is generically a complex number. In the present case the generators of the Kohno-Drinfield Lie algebra should be hermitian in order to have a unitary representation. Hence, we should have $\lambda(\theta) \in \mathbb{R}$. One can encompass all the UV $tt^{*}$ linear problems of the subbundles $\mathcal{V}_{N}^{k_{1},...,k_{d}}$ in a unique differential equation

\begin{equation}
\mathcal{D}  \mathbf{\Psi}=0, \hspace{1cm}   \mathbf{\Psi}\in \Gamma\left( C_{d}, \mathcal{V}^{k_{1},..., k_{d}}\right) 
\end{equation}

where $\Gamma\left( C_{d}, \mathcal{V}^{k_{1},...,k_{d}}\right) $ is the space of sections of the $(k_{1},...,k_{d})$-Gran bundle. The above equation is known as the Knizhnik-Zamolodchikov equation and was first introduced in \cite{wzw} as the isomonodromic PDE satisfied by the $d$-point correlators of the WZW conformal field theory. The holonomy of the KZ connection defines a one parameter family of linear representations $\rho_{\lambda}$ of the pure braid group $P_{d}=\pi_{1}(C_{d})$ of $d$ ordered distinct points 

\begin{equation}
\rho_{\lambda}: P_{d}\rightarrow \mathrm{End}\left( V^{k_{1}}\otimes \cdot\cdot\cdot \otimes V^{k_{d}}\right).
\end{equation}

In the case in which the punctures are all equal, namely $V^{k_{1}}=\cdot\cdot\cdot=V^{k_{d}}=V$, the KZ connection is invariant under the action of the symmetric group $S_{d}$ and descends naturally on the configuration space of $d$ unordered points $Y_{d}=C_{d}/S_{d}$. Thus, we have a one parameter family of linear representations of the Artin braid group $\mathcal{B}_{d}=\pi_{1}(Y_{d})$ of $d$ strings

\begin{equation}
\rho_{\lambda}: \mathcal{B}_{d}\rightarrow \mathrm{End}\left( V^{\otimes d}\right).
\end{equation}

In the case in which $V$ is the fundamental representation of $SU(2)$, $\rho_{\lambda}$ is a Hecke algebra representation of the braid group $\mathcal{B}_{d}$ which factors through the Temperley-Lieb algebra $A_{d}(q)$ with $q=e^{i\pi \lambda}$ \cite{hecke1,hecke2}.\\ The parameter $\lambda$ has been extimated in \cite{paper} with consistency arguments and turns out to be the following piecewise linear function of the $\theta$ angle

\begin{equation}
\lambda= \frac{\theta}{\pi} \ \mathrm{mod} \ 1= \frac{1}{\nu}  \ \mathrm{mod} \ 1,
\end{equation}

where we have used the relation \ref{nutheta}. If we want our $\mathcal{N}=4$ theory to admit an effective IR description in terms of a unitary $3d$ topological field theory, it is natural to consider the subset of solutions 

\begin{equation}
\lambda= \pm \frac{2}{\kappa+2} \ \mathrm{mod} \ 2, \hspace{1cm} \kappa \in \mathbb{Z}.
\end{equation}

In other words, we ask the UV Berry connection of the conformal blocks bundle to be a KZ connection for the $SU(2)$ current algebra of the WZW model with quantized level $\kappa$. This is known to be the theory of edge modes of the $SU(2)$ Chern-Simons theory with boundary. Comparing the above equations one finds the allowed values of the filling fraction given in \ref{all1} and \ref{all2}.

\subsection{The Preferred Ground state}\label{dictionary}

The KZ connection describes the non-abelian braiding properties of the quasi-holes operators $\Sigma^{k_{j}}_{j,m_{j}}(w_{j})$. The representation of the pure braid group of $d$ strings provided by the holonomy of the connection is highly reducible. The subbundles $\mathcal{V}_{N}^{k_{1},...,k_{d}}$  are eigenbundles of $L_{3}$ with eigenvalues

\begin{equation}
m=N-K/2
\end{equation}

and by construction must be preserved by parallel transport with the $tt^{*}$ connection. Consistently, the operators $\Omega_{ij}$ commute with $L_{3}=\hat{N}- K/2$ and the eigensubbundles $\mathcal{V}_{N}^{k_{1},...,k_{d}}$ are preserved by parallel transport with $\mathcal{D}$. We note that, since we must have $1\leq k_{j}\leq N$ for the $N$ electron sector, the eigenbundles with particle number $N$ smaller than some of the $k_{j}$ do not correspond to any Vafa LG model. However, since they are invariant subbundles and their states do not mix with the physical states, we can include them in the sum \ref{sum} without any harm.\\ The monodromy representation is further reducible, since also the operator $L^{2}= \sum_{\ell}L_{\ell}L_{\ell}$ is preserved by parallel transport. Thus, the KZ connection preserves all the eigenbundles $\mathcal{V}_{l,m}^{k_{1},..., k_{d}}$ of given total angular momentum 

\begin{equation}
\begin{split}
\psi \in &  \ \mathcal{V}_{l,m}^{k_{1},...,k_{d}}   \ \ \  \Longleftrightarrow \ \ \ (L^{2}-l(l+1))\psi= (L_{3}-m)\psi=0, \\ 
& \ m=-l,-l+1,.....,l-1,l  \ \ \ \ \ \ l \in \frac{1}{2}\mathbb{N}.
\end{split}
\end{equation}

Moreover, since the monodromy representation centralizes with respect to $L^{2}$, we have 

\begin{equation}\label{isoml}
\mathcal{V}_{l,m}^{k_{1},..., k_{d}} \simeq \mathcal{V}_{l,m^{\prime}}^{k_{1},..., k_{d}} , \ \ \ \ \ \mathrm{for} \ -l\leq m, m^{\prime} \leq l.
\end{equation}

The rank of the eigenbundle $\mathcal{V}_{l,m}^{k_{1},...,k_{d}} $ counts how many times the $SU(2)$ representation with spin $l$ satisfying

\begin{equation}
\vert m \vert \leq l \leq K/2
\end{equation}

appears in the decomposition of $V^{k_{1}}\otimes \cdot\cdot\cdot \otimes V^{k_{d}}$. In particular, the representation with highest angular momentum appears once in the tensor product, implying that each $N$-particle subbundle contains an eigenbundle $\mathcal{V}_{K/2, \, N- K/2}^{k_{1},...,k_{d}}$ of rank $1$, i.e. spanned by a unique monodromy invariant vacuum. This preferred state is identified in \cite{paper} with the topological vacuum of FQHE. By the \ref{isoml} we get 

\begin{equation}
\mathcal{V}_{K/2,\,N- K/2}^{k_{1},..., k_{d}}\simeq \mathcal{V}_{K/2, \, K/2}^{k_{1},...,k_{d}},
\end{equation}

where $\mathcal{V}_{K/2, \, K/2}^{k_{1},..., k_{d}} $ is the subbundle describing the $\nu=1$ phase. We also observe that, since $m$ and $-m$ appear with the same multiplicity in the decomposition of $V^{k_{1}}\otimes \cdot\cdot\cdot \otimes V^{k_{d}}$, we should have 

\begin{equation}
\mathcal{V}_{N}^{k_{1},...,k_{d}}\simeq \mathcal{V}_{K-N}^{k_{1},...,k_{d}}.
\end{equation}

So far we have discussed the braid representation of $\mathcal{Q}$-conformal blocks. It is argued in \cite{paper} that the natural connection on the $tt^{*}$ vacuum bundle differs from the connection $\mathcal{D}$ acting on the $\mathcal{Q}$-conformal blocks by a line bundle twisting. In particular, the horizontal sections of the KZ connection differ from the actual $tt^{*}$ brane amplitudes by a normalization factor given by some multivalued holomorphic function on $C_{d}$. In the case of minimal quasi-hole operators the normalization factor is given by the correlator $\langle \tau_{1}(w_{1})\cdot\cdot \cdot \tau_{d}(w_{d})\rangle$, which is a section of the line bundle $\mathcal{V}_{d/2,-d/2}^{1,..., 1}$. In the general case we can obtain the $tt^{*}$ connection acting on the vacua of the LG model by replacing the KZ connection with 

\begin{equation}
\mathcal{D}= d + \lambda\sum_{i<j} \frac{\Omega_{ij}}{w_{i}-w_{j}}d(w_{i}-w_{j}) + \sum_{i<j}\xi_{ij}d\log(w_{i}-w_{j})
\end{equation}

for some constants $\xi_{ij}$. Following the argument of \cite{paper}, these parameters should be determined such that the topological vacuum of FQHE has trivial monodromy for any $N$-particle sector. This condition fixes $\xi_{ij}$ in terms of $\lambda$ as 

\begin{equation}
\xi_{ij}=-k_{i}k_{j} \frac{\lambda}{4}.
\end{equation}

In other words, the normalized brane amplitudes $\mathbf{\Psi}_{tt^{*}}$ are related to the KZ ones $\mathbf{\Psi}$ by

\begin{equation}
\mathbf{\Psi}_{tt^{*}}= \frac{\mathbf{\Psi}}{\Psi_{\mathrm{priv}}},
\end{equation}

where $\Psi_{\mathrm{priv}}$ is a parallel section of the line bundle $\mathcal{V}_{K/2, \, K/2}^{k_{1},..., k_{d}} $. We note that the normalized monodromy is trivial in the $N=K$ sector, consistently with the fact the $\nu=1$ phase has trivial topological order.

\section{FQHE on Riemann Surfaces}\label{riemann}

In this section we briefly discuss the higher genus generalization of the Vafa model of FQHE and the corresponding $tt^{*}$ geometry. 

\subsection{Lowest Landau Level on Riemann Surfaces}

The Hamiltonian describing the dynamics of a bunch of $N$ electrons in a constant uniform magnetic field $B$ on a generic Riemann surface may be written as sum of two pieces 

\begin{equation}
H= H_{B}+ H_{\mathrm{int}},
\end{equation}

where $H_{B}$ describes the interaction between the electrons and the magnetic field $B$, while $H_{\mathrm{int}}$ contains all the other possible interactions, with the crucial property that it is $O(1)$ as $B\rightarrow \infty$ (see \cite{wen,tong}). We denote with $\mathbf{H}_{\Phi}\subset \mathbf{H}$ the subspace of the Hilbert space containing states whose energy is bounded in the limit of strong magnetic field $B$. Denoting with $\Phi >0$ the magnetic flux of $B$ through the Riemann surface, the dimension of $\mathbf{H}_{\Phi}$ is given by 

\begin{equation}
\mathrm{dim} \ \mathbf{H}_{\Phi}=  \begin{pmatrix} \Phi/2\pi \\ N \end{pmatrix}
\end{equation}

where we assume 

\begin{equation}
N=\nu \frac{\Phi}{2\pi}, \hspace{1cm}  \ 0<\nu\leq 1, \ \ \nu \in \mathbb{Q}.
\end{equation}

The orthogonal complement in the full Hilbert space is separated from $\mathbf{H}_{\Phi}$ by a large $O(B)$ energy gap. Moreover, in $\mathbf{H}_{\Phi}$ the electrons are polarized and the spin degrees of freedom are frozen. Thus, if we are interested in an effective description at energies $\ll B$\footnote{We set the mass of the electron $m_{e}$ to $1$.}, we are reduced to a quantum system with finite dimensional Hilbert space $\mathbf{H}_{\Phi}$ and Hamiltonian 

\begin{equation}
H_{\mathrm{eff}}=P_{\Phi} H P_{\Phi},
\end{equation}

where $P_{\Phi}$ is the projector on $\mathbf{H}_{\Phi}$.\\ 
 Since the Landau levels are all isomorphic to each other, we may assume without loss of generality that the electrons fill a fraction of the lowest Landau level (see \cite{tong}). Following \cite{paper}, one can describe the lowest Landau level of a single electron on a Riemann surface $\Sigma$ of arbitrary genus $g$ using the language of algebraic geometry. We consider a complex line bundle $\mathcal{L}\rightarrow \Sigma $ with first Chern class $c_{1}(\mathcal{L})= \Phi/2\pi$. The complex structure of the Riemann surface is irrelevant in the discussion and we can choose it according to convenience. A state of the system is represented by a smooth section $\psi: \Sigma\rightarrow \mathcal{L}$ of the line bundle. Every complex line bundle over a one dimensional complex manifold is holomorphic and so admits holomorphic trivializations. The line bundle is endowed with an hermitian metric $h$ and the inner product on the Hilbert space is defined as

\begin{equation}
\langle \psi_{1} \vert \psi_{2} \rangle = \int_{\Sigma} dzd\bar{z} h \ \psi_{1}^{*}\psi_{2}.
\end{equation}

The lowest Landau level is defined as the kernel of the Laplacian

\begin{equation}
\nabla=-D_{z}D_{\bar{z}} 
\end{equation}

where $D$ is the $U(1)$ Chern connection associated to $h$ (see \cite{wen,thesis}). In an holomorphic trivialization one has 

\begin{equation}
D_{z}=-h\partial_{z}h^{-1}, \hspace{1cm} D_{\bar{z}}=\partial_{\bar{z}}.
\end{equation}

The magnetic field is the curvature of the connection, which is a closed and real $(1,1)$ form 

\begin{equation}
F_{h}=\bar{\partial}\partial \log h,
\end{equation}

where $\partial, \bar{\partial}$ are the Dolbeault operators of the complex manifold. The Riemann surface $\Sigma$ endowed with the curvature of the line bundle is naturally a Kahler manifold with a globally defined Kahler potential $K=\log h$. Isomorphism classes of line bundles over a compact Riemann surface are in correspondence modulo linear equivalence with the divisors of the surface 

\begin{equation}
D= \sum_{i=1}^{d} n_{i} p_{i},
\end{equation}

where $p_{i} \in \Sigma$ and $n_{i}$ are integers. From the physical point of view, $D$ describes the magnetic background in which the electron moves. We demand this divisor to be effective, namely with $n_{i}>0$. Since the electrons are particles with spin, in this discussion we have also to endow the line bundle with a spin structure. This is associated with a divisor $S$ on the Riemann surface such that $2S=K$, where $K$ is in the canonical divisor of $\Sigma$. The divisor identifying the twisted line bundle is the sum $D+S$ and is unique up to linear equivalence. The vacuum wave functions satisfy $D_{\bar{z}} \phi=0$ and so we have the definition 

\begin{equation}
\mathbf{H}_{\Phi,1 \, \mathrm{part.}}= \Gamma (\Sigma, \mathcal{O}(D+S)), \hspace{1cm} \mathrm{dim} \ \mathbf{H}_{\Phi,1 \, \mathrm{part.}}= \mathrm{deg} \ \mathcal{L}(D+S)= \frac{\Phi}{2\pi},
\end{equation}

where $\mathbf{H}_{\Phi,1 \, \mathrm{part.}}$ is the lowest Landau level and $\Gamma (\Sigma, \mathcal{O}(D+S))$ is the space of holomorphic sections of the twisted line bundle $\mathcal{L}(D+S)$. It is convenient to pick an even spin structure with $\mathrm{dim} \ \Gamma (\Sigma, \mathcal{L}(S))=0$ associated to a divisor $S=\sum_{\ell=1}^{g}r_{\ell}-q$, where $r_{\ell},q$ are distinct points on the Riemann surface which satisfy $\sum_{\ell=1}^{g} 2 r_{\ell}-K=2q$. The choice of $r_{\ell}$ does not affect the discussion and we can translate them as we prefer. The divisor $D+S$ has a defining meromorphic section $\psi_{0}$ with zeros of order $n_{i}$ at $p_{i}$, simple zeros at $r_{\ell}$ and a simple pole at $q$. The map $\psi\rightarrow \psi/\psi_{0}=\phi$ provides a canonical identification  

\be
\begin{split}
\mathbf{H}_{\Phi,1 \, \mathrm{part.}}&\equiv\Gamma\!\big(E,\co(D+S)\big)\xrightarrow{\!\sim\!}\\
&\xrightarrow{\!\sim\!}
\Big\{\text{$\phi\in\Gamma(\Sigma,\mathscr{M})$ with polar divisor $D_\infty\leq D+\sum_{\ell}r_\ell$ vanishing at $q$}\Big\},
\end{split}
\ee

where $\Gamma(\Sigma,\mathscr{M})$ is the space of meromorphic functions on $\Sigma$. Then, let $z_i$ be a local parameter at $p_i\in \Sigma$ and $P\!P_p(\phi)$ the principal of the meromorphic function $\phi$ at $p\in \Sigma$. The map 

\be
\phi\longmapsto \Big\{ z_1^{n_1}\, P\!P_{p_1}(\phi), z_2^{n_2}\, P\!P_{p_2}(\phi),\cdots, z_d^{n_d}\, P\!P_{p_d}(\phi)\Big\} 
\ee

defines a linear isomorphism between the lowest Landaul level and the complex algebra

\be\label{lllasqw}
\mathbf{H}_{\Phi,1 \, \mathrm{part.}}\xrightarrow{\!\sim\!} \prod_{i=1}^ d  \C[z]\big/(z^{n_i}). 
\ee

In order to get the Hilbert space $\mathbf{H}_{\Phi}$ of the $N$-electron system we simply need to take the antisymmetric tensor product of the single particle space

\begin{equation}
\mathbf{H}_{\Phi}= \bigwedge^{N} \Gamma (\Sigma, \mathcal{L}(D+S)), \hspace{1.5cm} \mathrm{dim} \ \mathbf{H}_{\Phi}= \begin{pmatrix} \Phi/2\pi \\ N \end{pmatrix}.
\end{equation}

It is manifest by \ref{lllasqw} that the Hilbert spaces $\mathbf{H}_{\Phi} $ associated to Riemann surfaces of different genera are all isomorphic to each other.

\subsection{$\mathcal{N}=4$ Supersymmetric Description}

Generalizing the proof of \cite{paper}, one can show that the effective Hilbert space $\mathbf{H}_{\Phi}$ is isomorphic to the vacuum space of a certain $\mathcal{N}=4$ supersymmetric model. A theory in supersymmetric quantum mechanics with four supercharges is specified by the choice of a Kahler potential $K$ and an holomorphic superpotential $W$ \cite{sqm}. Once the topology of the Riemann surface is fixed, the choice of the Kahler metric does not affect the structure of the vacua, which depends only on the superpotential. Hence, we can choose $K=\log h$ as globally defined Kahler potential on the Riemann surface. The number of vacua is given by the Witten index, which is equal to the number of zeros of the $1$-form $dW$ counted with multiplicitly \cite{sqm}. In order to compare this description with the previous one we choose as target space the manifold $\mathcal{K}=\Sigma\setminus \mathrm{supp}\ F$, where $F$ is an effective divisor on $\Sigma$. As we discussed above, a divisor identifies up to linear equivalence a line bundle with its set of holomorphic sections. However, for a linear combination of points on the Riemann surface one can associate also a $\mathcal{N}=4$ supersymmetric system. Given an effective divisor $D=\sum_{i}n_{i}p_{i}$, we assign a closed meromorphic $1$-form $dW$ on $\Sigma$ with zeros in $p_{i}$ of order $n_{i}$. The polar divisor of $W^{\prime}(z)$ is given by $F\sim D$. To make precise the correspondence between the two descriptions, we set $r_\ell \in \mathrm{supp}\ F$ so that $r_\ell \not\in \mathcal{K}$. By consistency we should have $\mathrm{deg} (F) = \mathrm{deg} (D) > g$. The chiral ring $\mathcal{R}$ of the LG model is defined as \cite{sqm}

\begin{equation}
 \mathcal{R}=\Gamma(\mathcal{K},\mathscr{O}_{\mathcal{K}})/\Gamma(\mathcal{K},\mathcal{J}_{W}),
\end{equation}

where $\mathscr{O}_{\mathcal{K}}$ is the space of holomorphic functions on $\mathcal{K}$ and $ \mathcal{J}_{W}\subset \mathscr{O}_{\mathcal{K}}$ is the Jacobian ideal generated by $\partial_{z} W$. By the chinese remainder theory we have 

\begin{equation}
\mathcal{R} \cong \prod_{i=1}^ d  \C[z]\big/(z^{n_i}).
\end{equation}

Comparing with \ref{lllasqw} we get an isomorphism of vector spaces

\be
\mathbf{H}_{\Phi,1 \, \mathrm{part.}}\cong \mathcal{R},\hspace{1.5cm} \mathbf{H}_{\Phi}\cong\bigwedge\nolimits^{\!N}\mathcal{R}.
\ee

An explicit realization is provided by the map 

\begin{equation}\label{is1}
\frac{\psi}{\psi_{0}}\rightarrow \frac{\psi}{\psi_{0}}W^{\prime} = \mathcal{O} \in \mathcal{R}.
\end{equation}

On the other hand, we also have a linear isomorphism between $\mathcal{R}$ and the space of SUSY vacua $V$ \cite{sqm}. The ground states can be written in Schroedinger representation as $L^{2}$-differential forms on $\mathcal{K}$  

\begin{equation}\label{is2}
\psi_{\mathrm{SUSY}}= \mathcal{O} dz + \overline{\mathcal{Q}}(....),
\end{equation}

where $\overline{\mathcal{Q}}= \bar{\partial} + dW \wedge$ is a SUSY charge. The vacuum space is isomorphic to the $\overline{\mathcal{Q}}$-cohomology with $L^{2}$ coefficients, whose classes are labelled by the chiral operators $\mathcal{O} \in \mathcal{R}$. Composing \ref{is1} and \ref{is2} we have an isomorphism between the low-lying states of the two quantum systems

\begin{equation}
\mathbf{H}_{\Phi,1 \, \mathrm{part.}} \xrightarrow{\sim} V, \hspace{1.5 cm} \mathbf{H}_{\Phi} \xrightarrow{\sim} V_{\Phi, N}= \bigwedge^{N} V.
\end{equation}

\subsection{The Vafa Superpotential}

Under the isomorphism constructed above, the effective Hamiltonian $H_{\mathrm{eff}}=P_{\Phi} H P_{\Phi}$ is mapped to some Hamiltonian 

\begin{equation}\label{ham}
\tilde{H}= H_{\mathcal{W}} + H_{\mathrm{su.br}}
\end{equation}

where $H_{\mathcal{W}} $ is a $\mathcal{N}=4$ supersymmetric Hamiltonian and $H_{\mathrm{su.br}}$ contains the SUSY breaking interactions. Let us first discuss the supersymmetric part. The choice of the Kahler potential has no influence on the IR dynamics of the quantum system. What is relevant for monodromy considerations is the $1$-form $d\mathcal{W}$. This should be a meromorphic differential on $\mathcal{K}^{N}$ invariant under permutation of the electron coordinates. Under the projection $\mathcal{K}^{N}\rightarrow \mathcal{K}_{i}$ on the target space of the $i$-th electron, $d\mathcal{W}$ reduces to a meromorphic $1$-form $dU$ with polar divisor 

\begin{equation}
\tilde{F}= F + \sum_{j\neq i} n_{j}q_{j}, \hspace{1cm} n_{j}>0,
\end{equation}

where $q_{j}$ denote the fixed positions of the other electrons. We choose a divisor $F$ of the form

\begin{equation}
F=\sum_{k=1}^{\Phi/2\pi-h} \zeta_{k} + \sum_{a=1}^{h} x_{a}.
\end{equation}

We have two types of topological defects. The $\zeta_{k}$ form a regular distribution on $\Sigma$ which mimics the constant macroscopically uniform magnetic field. The compatibility with the Hermitian structure of $\mathbf{H}_{\Phi}$ requires to choose the residues of $dU$ at the $\zeta_{k}$ to be $\pm 1$, where the two possibilities are related by a change of orientation \cite{paper}. Here we fix the conventions such that the residues are equal to $-1$. A small number $h$ of quasi-holes are inserted on $\mathcal{K}$ at positions $x_{a}$. These are modelled by simple poles of residue $+1$. As we introduce the quasi-holes in the system, there is a small mismatch between the dimension of the lowest Landau level and the effective magnetic flux. The differential $dU$ describes also the supersymmetric part of the interaction between the $i$-th electron and the other ones. An electron of fixed position is equivalent to a hole. Thus, $dU$ should contain simple poles at $q_{j}$ (i.e. $n_{j}=1$) with equal residues by permutation symmetry.\\ We have enough informations to constraint the form of $d\mathcal{W}$. Let $E(p,q)$ be the prime form of the Riemann surface for two points $p,q \in \mathcal{K}$. Then let $z_{i},i=1,...,N$ be the local coordinates of the electrons on the target manifold $\mathcal{K}^{N}$. The $1$-form $d\mathcal{W}$ which defines the higher genus generalization of the Vafa LG model should be a meromorphic differential of the form 

\begin{equation}
d\mathcal{W}=\sum_{i=1}^{N}\left(\mu+ \sum_{a}U(z_{i},x_{a})-\sum_{k}U(z_{i},\zeta_{k}) + 2\beta\sum_{j\neq i} 
U(z_{i},z_{j})\right)dz_{i} ,
\end{equation}

where 

\begin{equation}
U(z,w)= \frac{E^{\prime}(z,w)}{E(z,w)}.
\end{equation}

As function of the position $z_{i}$ of the $i$-th electron, the meromorphic function 

\begin{equation}
U(z_{i};x_{a},\zeta_{k},z_{j})=\mu+ \sum_{a}U(z_{i},x_{a})-\sum_{k}U(z_{i},\zeta_{k}) + 2\beta\sum_{j\neq i}U(z_{i},z_{j})
\end{equation}

has the form of $2d$ electrostatic field of a system of point charges on $\Sigma$ superimposed to a constant background field $\mu$. This coupling is an integration constant which regularizes the superpotential at infinity in the planar case. On compact Riemann surfaces of genus $g\geq 1$ we do not have escaping vacua at infinity in the limit $\mu\rightarrow 0$. Thus, we could simply take $\mu=0$ without affecting the Witten index of the theory and the monodromy representation. Despite this, since the $tt^{*}$ monodromy representation does not depend on the choice of $\mu$ and we want to recover the properly regularized superpotential as we take the planar limit, we keep $\mu \in \mathbb{C}^{\times}$. Since $U(z_{i};x_{a},\zeta_{k},z_{j})$ should be a meromorphic function on $\mathcal{K}$, the size of the Coulomb repulsion $\beta$ must be fixed such that the sum of residues vanishes. We have 

\begin{equation}
0= -\left( \frac{\Phi}{2\pi}-h \right)  + h + 2\beta (N-1)\approx (2\beta \nu-1)\frac{\Phi}{2\pi},
\end{equation}

where in the last equality we have used $N,\Phi \gg 1$ and the relation $N=\nu \, \Phi/2\pi$. Hence, we obtain the quantization condition 

\begin{equation}
2\beta= 1/\nu \in \mathbb{Q}_{>0}
\end{equation}

which is the coupling given in \cite{vafa}. We observe that, differently form the genus $0$ case, once we fixed the residues of the poles of $d\mathcal{W}$ for a given number $d=\Phi/2\pi$ of topological defects and filling fraction $\nu$, the Vafa Hamiltonian is defined on Riemann surfaces of higher genus only for $N=\nu d$.

\subsection{Topological Degeneracy}

The Hamiltonian $H_{\mathrm{su.br.}}$ contains the SUSY breaking interactions of the full Hamiltonian $\tilde{H}$. While the Vafa Hamiltonian is of order $O(B)$, the SUSY breaking piece is of order $O(1)$ and so a small pertutrbation for large magnetic fields. However, we are not allowed to neglect it when we study the topological order of FQHE. Indeed, $H_{\mathrm{su.br.}}$ should lift the huge deneracy of the ground state of $H_{\mathcal{W}}$ leaving only the topological vacuum. Let us first recall the situation in the case of the complex plane. The fermionic zero energy states of the Vafa Hamiltonian with $N$ electrons and $d$ punctures generate a vacuum bundle over the space of couplings $C_{d}$ entering in the superpotential 

\begin{equation}
\mathcal{V}_{d,N}\rightarrow C_{d}, \hspace{1cm} \mathrm{rank} \ \mathcal{V}_{d,N}= \begin{pmatrix} d \\ N \end{pmatrix}, \hspace{1cm} N= \nu d.
\end{equation}

The vacuum bundle is equipped with the $tt^{*}$ Berry connection $\mathcal{D}=d+A$, which describes the unitary time evolution of the quantum system. The $tt^{*}$ equations in the UV limit of the $\mathcal{N}=(2,2)$ version of the model requires $\mathcal{D}$ to be meromorphic and flat, i.e.

\begin{equation}\label{uveqn}
\bar{\partial} A=0, \hspace{1cm} \mathcal{D}^{2}=0.
\end{equation}

The UV Berry connection provides the ground state $V_{d,N}$ with the structure of a unitary monodromy representation of $C_{d}$ 

\begin{equation}
\rho_{\nu,d}: \pi_{1}(C_{d}) \rightarrow \mathrm{End} (V_{d,N})
\end{equation}

which captures the quantum topological order of $H_{\mathcal{W}}$. As we switch on the SUSY breaking Hamiltonian, the degeneracy of zero energy states is lifted and the ground state decomposes into eigenspaces of the full Hamiltonian $\tilde{H}$. In particular, $H_{\mathrm{su.br.}}$ selects a unique state which corresponds to the topological vacuum of FQHE on the complex plane. As argued in \cite{paper}, as long as $H_{\mathrm{su.br.}}$ is invariant under permutations of electrons and topological defects of equal charge, its eigenspaces must be subrepresentations of $\rho_{\nu,d}(\pi_{1}(C_{d}))$. Therefore, the $tt^{*}$ connection $\mathcal{D}$ can capture the topological order of the full Hamiltonian as well. By consistency, $\rho_{\nu,d}$ must be reducible with in particular an invariant subbundle of rank $1$. The fiber of this line bundle is spanned by the maximally symmetric state of $V_{d,N}$ with respect to arbitrary permutations of the punctures. Such preferred ground state corresponds to the unique topological vacuum of FQHE on the plane and its existence has been cheked in \cite{paper}. Since the assumptions on $H_{\mathrm{su.br.}}$ are independent from the details of the interactions, the conclusion of the argument is that the Vafa $\mathcal{N}=4$ SUSY Hamiltonian represents a topological universality class which contains any Hamiltonian describing the motion of electrons in a strong magnetic field.\\ Analogously to the Vafa Hamiltonian, any representative $H_{\mathrm{su.br.}}$ of the universality class should admit an higher genus generalization. Besides preserving the indistinguishability of the electrons, the necessary condition that the full Hamiltonian $\tilde{H}$ should satisfy to have a $4$-SUSY uplift is the following: its space of couplings should match the $tt^{*}$ one \cite{paper}. This is in general a very restrictive property, since the $tt^{*}$ space of couplings must be a complex analytic space with globally defined Kahler potential and a Frobenious structure \cite{dubr}. Thus, it is very remarkable that the space of essential parameters of FQHE satisfies these peculiar conditions. The manifold of couplings entering in $\mathcal{W}$ is the moduli space of complex structures $\mathcal{P}_{g,d}$ of a Riemann surface $\Sigma_{g,d}$ of genus $g$ with $d$ punctures. This space has the structure of an orbifold 

\begin{equation}
\mathcal{P}_{g,d}= \mathcal{T}_{g,d}/\mathcal{M}_{g,d}
\end{equation}

where $\mathcal{T}_{g,d}$ is the Teichmuller space and the orbifold fundamental group $\mathcal{M}_{g,d}$ is the mapping class group (see \cite{teich} for an introduction). A point of $\mathcal{T}_{g,d}$ identifies an isomorphism class of Riemann surfaces $\Sigma_{g,d}$ related by an holomorphic diffeomorphism $f \in \mathrm{Diff}_{0}\left( \Sigma_{g,d}\right) $ isotopic to the identity. The mapping class group is defined as the group of isotopy classes of orientation-preserving diffeomorphisms

\begin{equation}
\mathcal{M}_{g,d}= \mathrm{Diff}^{+}\left( \Sigma_{g,d} \right) /\mathrm{Diff}_{0}\left( \Sigma_{g,d}\right)  =\pi_{0}\left( \mathrm{Diff}^{+}\left( \Sigma_{g,d} \right) \right), 
\end{equation}

where $\mathrm{Diff}^{+}\left( \Sigma_{g,d} \right)$ is the space of holomorphic diffeomorphisms which preserve the orientation. We have a chain of canonical inclusions

\begin{equation}
\mathcal{M}_{0,d} \xrightarrow{\iota_{0,d}} \mathcal{M}_{1,d} \xrightarrow{\iota_{1,d}}\cdot \cdot \cdot \xrightarrow{\iota_{g-1,d}} \mathcal{M}_{g,d} \xrightarrow{\iota_{g,d}} \mathcal{M}_{g+1,d} \xrightarrow{\iota_{g+1,d}} \cdot \cdot\cdot  
\end{equation}

where $\iota_{g,d}$ are injective homomorphisms. If we want to classify punctured Riemann surfaces $\Sigma_{g,d}$ up to strict equivalence we should consider $\mathcal{P}_{g,d}$ as the actual manifold of couplings. This space is described at genus $g\geq 2$ by $3g-3+d$ complex parameters, while in the case of the torus we have $d+1$ moduli. In the planar case the moduli space of complex structures is just the configuration space of $d$ ordered points on $\mathbb{C}$ modulo permutations of the ones with equal charge.\\ 
The topological universality class represented by the $N$-particle Vafa Hamiltonian is a functor which associates to any Riemann surface $\Sigma_{g,d}$ an Hilbert space $V_{d,N}$ endowed with the structure of a unitary monodromy representation of $\mathcal{M}_{g,d}$ 

\begin{equation}\label{towerrep}
 \rho_{\nu,g,d}: \mathcal{M}_{g,d}\rightarrow \mathrm{End}(V_{d,N}). 
\end{equation}

These representations are generated by parallel transport with the $tt^{*}$ flat connection and satisfy the functorial induction and restriction relations

\begin{equation}
\rho_{\nu,g,d}( \mathcal{M}_{g,d}) \xrightarrow{\iota^{*}_{g,d}} \rho_{\nu,g+1,d}( \mathcal{M}_{g+1,d}), \hspace{1cm} \rho_{\nu,g+1,d}( \mathcal{M}_{g,d})= \rho_{\nu,g,d}( \mathcal{M}_{g,d}), 
\end{equation}

where $\iota^{*}_{g,d}$ is an injective homomorphism induced by $\iota_{g,d}$. The class of monodromy representations \ref{towerrep} defines the underlying $3d$ topological quantum field theory of the Vafa universality class. The analysis of \cite{paper} for the $g=0$ case shows that the topological order of $H_{\mathcal{W}}$ for a given $\nu$ is captured by the $SU(2)$ Chern-Simons theory at the corresponding integral level $ \kappa$. Thus, the Vafa LG model must define the same tower of monodromy representations. Any topological defect in the spectrum of the $SU(2)$ Chern-Simons theory carries some spin $ k/2 \in \frac{1}{2}\mathbb{N}$ degree of freedom and is associated to the corresponding representation $V^{k}\simeq \mathbb{C}^{k+1}$ of $su(2)$. We can generalize the construction of section \ref{spindeg} and define the Grand bundle of $\mathcal{Q}$-conformal blocks over moduli spaces

\begin{equation}
\mathcal{V}^{k_{1},...,k_{d}}_{g,d}\rightarrow \mathcal{P}_{g,d}, \hspace{1cm} \mathrm{rank} \ \mathcal{V}^{k_{1},...,k_{d}}_{g,d}= \prod_{j=1}^{d}(k_{j}+1)
\end{equation}

with fiber the tensor product space 

\begin{equation}
V^{k_{1},...,k_{d}}= V^{k_{1}}\otimes \cdot\cdot\cdot \otimes V^{k_{d}}.
\end{equation}

The Gran bundles $\mathcal{V}^{k_{1},...,k_{d}}_{g,d}$ are equipped with the higher genus generalization of the Knizhnik-Zamolodchikov connection, also known as the Knizhnik-Zamolodchikov-Bernard connection. The detailed construction of this connection on generic Riemann surfaces can be found in \cite{kzb}. The non-normalized $tt^{*}$ brane amplitudes $ \mathbf{\Psi}$ form in the UV limit a local basis of holomorphic sections of $\mathcal{V}^{k_{1},...,k_{d}}_{g,d}$ satisfying the Knizhnik-Zamolodchikov-Bernard equations, which we can write schematically as 

\begin{equation}\label{system}
\left( \partial_{w_{j}} + \lambda H_{j}\right) \mathbf{\Psi}=0 , \hspace{1cm} \left( \partial_{\tau_{\mu}} + \lambda H_{\mu}\right) \mathbf{\Psi}=0, \hspace{1cm} \lambda= \pm \frac{2}{\kappa+2}, \ \ \ \kappa \in \mathbb{Z},
\end{equation}

where $w_{j}, j=1,...,d$ denote the position of the punctures $\zeta_{k},x_{a}$ and $\tau_{\mu}$ are the extra moduli on $\mathcal{P}_{g,d}$ ($\mu=1,...,3g-3$ for $g\geq 2$ and $\mu=1$ for $g=1$). In genus $0$ the couplings are just the $w_{j}$ and the operators $H_{j}$ reduce to the Gaudin Hamiltonians entering in the KZ equations

\begin{equation}
H_{j}^{g=0}= \sum_{i\neq j} \frac{s_{i}^{\ell}s_{j}^{\ell}}{w_{i}-w_{j}}.
\end{equation}

The components of the connection satisfy the integrability conditions

\begin{equation}
\left[ \partial_{j}+\lambda H_{j}, \partial_{k}+\lambda H_{k}\right] = \left[ \partial_{j}+\lambda H_{j}, \partial_{\mu}+\lambda H_{\mu}\right]= \left[ \partial_{\mu}+\lambda H_{\mu}, \partial_{\nu}+\lambda H_{\nu}\right]=0,   
\end{equation}

which guarantee the flatness of $\mathcal{D}$ as demanded by the UV $tt^{*}$ equations \ref{uveqn}. By parallel transport with the KZB connection, the vector spaces $V^{k_{1}}\otimes \cdot\cdot\cdot \otimes V^{k_{d}}$ form a collection of monodromy representations of $\mathcal{M}_{g,d}$. The KZB equations have been derived in \cite{bern1} for the torus and in \cite{bern2} for generic Riemann surfaces. They are known to be the system of isomonodromic PDEs satisfied by the conformal blocks of the $SU(2)$ WZW model at level $\kappa$. As argued in \cite{paper}, this CFT plays the role of $\mathcal{Q}$-theory of edge modes for the quantum Hall states.\\ The minimal topological defect is associated with the fundamental representation $V^{1} \simeq \mathbb{C}^{2}$ of $SU(2)$. As explained in section \ref{spindeg}, a quasi-hole in the spin up/down state corresponds to an occupied/empty single particle vacuum. In particular, the angular momentum $L_{3}$ is related to the particle number operator by $L_{3}=\hat{N}-d/2$. It is clear that the monodromy representation $\mathcal{V}^{1,...,1}_{g,d}$ must be reducible and any subbundle $\mathcal{V}^{1,...,1}_{g,d,N}$ with definite number of particles must be preserved by parallel transport with the $tt^{*}$ connection. The corresponding fiber  

\begin{equation}
V_{d,N} \subset V^{1}\otimes \cdot\cdot\cdot \otimes V^{1}, \hspace{1cm} \mathrm{dim} \ V_{d,N}= \begin{pmatrix} d \\ N \end{pmatrix}
\end{equation}

is an eigenspace of $L_{3}$ with eigenvalue $m=N-d/2$. We remark that only the eigenbundles with $N=\nu d$ for the allowed values of $\nu$ can play the role of vacuum bundle for some Vafa LG model. As we switch on the SUSY breaking interactions $H_{\mathrm{su.br}}$, the degeneracy of ground states of $V_{d,N}$ is partially lifted and we remain with the topological vacuum $V^{\mathrm{vac}}_{g,\nu} $ of FQHE at given $g$ and $\nu=N/d$. Since it is an uplift of the Vafa Hamiltonian, the topological order of $\tilde{H}$ must be captured by the $tt^{*}$ flat connection. It follows that the monodromy subrepresentation $\mathcal{V}^{1,...,1}_{g,d,N}$ of $\mathcal{D}$ must be further reducible, with a invariant subbundle $\mathcal{V}^{\mathrm{vac}}_{g,\nu}$ whose fiber corresponds to $V^{\mathrm{vac}}_{g,\nu}$. The rank of this preferred subbundle defines the topological degeneracy of the quantum Hall state as function of the filling fraction $\nu$ and the genus $g$ of the Riemann surface. This must be the same of the $SU(2)$ Chern-Simons theory and is computed by the Verlinde formula \cite{verlinde}

\begin{equation}
\mathrm{rank} \ \mathcal{V}^{\mathrm{vac}}_{g,\nu}= \left( \frac{\kappa+2}{2}\right) ^{g-1} \sum_{n=0}^{\kappa} \left( \sin \frac{(n+1)\pi}{\kappa+2}\right)^{2(1-g)} , \ \ \  g\geq 1,
\end{equation}

where the correspondence between the level $\kappa$ and the filling fraction $\nu$ is provided by \ref{all1} and \ref{all2}.

\section{Conclusions}

In this paper we extended the analysis made in \cite{paper} of the Vafa $\mathcal{N}=4$ SUSY model of FQHE. The Vafa Hamiltonian represents a topological universality class whose underlying topological quantum field theory is the $SU(2)$ Chern-Simons theory. This provides all the topological invariants which characterize the topological order of FQHE. The main examples are the braiding properties of anyonic particles, which are described by the Knizhnik-Zamolodchikov connection, and the topological degeneracy of the ground state which is given by the Verlinde formula. In principle the Vafa Hamiltonian should represent the topological universality class of Hamiltonians describing the dynamics of electrons coupled to a strong magnetic field. It would be interesting to verify if the experimentally observed quantum Hall states may be classified and described according to this construction.


\begin{thebibliography}{100}

\bibitem{vafa}
C. Vafa, ``Fractional Quantum Hall Effect and M-Theory '', arXiv:1511.03372 [cond-mat.mes-hall].

\bibitem{tong}
D.Tong, `` The Quantum Hall Effect '', http://www.dampt.cam.ac.uk/user/tong/qhe.html. 

\bibitem{wen}
X.G. Wen, `` Quantum field theory of many-body systems'',  Oxford graduate texts, 2004.

\bibitem{paper}
R.Bergamin and S.Cecotti, ``FQHE and $tt^{*}$ geometry'', arXiv:1910.05022 [hep-th], 2019

\bibitem{nab2}
G.W. Moore and N. Read, `` Nonabelions in the fractional quantum Hall effect '', Nucl. Phys. B 360, 362 (1991).

\bibitem{nab3}
N.Read and E. Rezay, `` Beyond paired quantum Hall States: Parafermions and incompressible states in the first excited Landau level '', Phys. Rev. B 59, 8084 (1999)[cond-mat/9809384].

\bibitem{rif10} 
S. Cecotti, C. Vafa, `` Topological-anti-topological fusion'', nuclear Physics B 367 (1991) 359-461.

\bibitem{rif12}
S.Cecotti, C.Vafa, `` On classification of N=2 Supersymmetric Theories '', hep-th/9209085, November 1992.

\bibitem{rif23}
S.Cecotti, A. Neitzke and C.Vafa ``Twistorial Topological Strings and a $tt^{*}$ Geometry for $\mathcal{N}=2$ Theories in $4d$ '' arXiv:1412.4793 [hep-th], 15 Dec 2014. 

\bibitem{thesis}
R.Bergamin, ``FQHE and $tt^{*}$ geometry'', arXiv:1910.07369 [cond-mat.mes-hall].

\bibitem{sqm}
S.Cecotti, L.Girardello, A.Pasquinucci, `` Singularity-Theory and $N=2$ Supersymmetry '' , Nucl. Phys. B328 (1989) 701; Int. J. Mod. Phys. A6 (1991) 2427

\bibitem{verlinde}
E.Verlinde, `` Fusion rules and  modular transformations in $2d$ conformal field theory'',  Nuclear Physics B300 [FS22] (1988) 360-376. 

\bibitem{bern1}
 D.Bernard, `` On the Wess-Zumino-Witten models on the torus'',  Nuclear Physics B303 (1988)77-93.

\bibitem{bern2}
D.Bernard, `` On the Wess-Zumino-Witten models on the Riemann surfaces'',  Nuclear Physics B309 (1988) 145-174

\bibitem{kzb}
D.Ivanov `` Knizhnik-Zamolodchikov-Bernard equations on Riemann surfaces'', hep-th/9410091.

\bibitem{morse} 
K.Hori, A. Iqbal and C.Vafa, ``D-branes and Mirror symmetry '', hep-th/0005247.

\bibitem{GW}
D. Gaiotto and E. Witten, ``Knot invariants from four-dimensional gauge theory'', arXiv:1106.4789.

\bibitem{ising}
S.Cecotti, C.Vafa, ``Ising Model and N=2 Supersymmetric Theories '', arXiv:hep-th/9209085.

\bibitem{kz}
T.Kohno , ``Homological Representations of Braid Groups and KZ Connections'', Journal of Singularities Volume 5 (2012), 94-108.

\bibitem{drinkon}
C.A.Abad,  `` Introduction to Representations of Braid Groups'', arXiv:1404.0724v1 (2014).

\bibitem{wzw}
V. G. Knizhnik and A. B. Zamolodchikov, `` Current algebra and Wess-Zumino models in two dimensions'', Nuclear Phys. B247 (1984), 83–103

\bibitem{hecke1}
T. Kohno, `` Monodromy representations of braid groups and Yang-Baxter equations'', Ann. Inst. Fourier, Grenoble 37, 4 (1987), 139-160.

\bibitem{hecke2}
T.Kohno, `` Hecke algebra representations of braid groups and classical Yang-Baxter equations'', in Conformal Field Theory and Solvable Lattice Models, Adv. Stud. in Pure Math. 16 (1988) pp. 255-269.

\bibitem{hecke3}
T. Kohno, `` Conformal field theory and topology'', Translations of Math. Monographs vol. 210, AMS (2002)

\bibitem{dubr}
B. Dubrovin, ``Geometry of 2d topological field theories'', arXiv:hep-th/9407018

\bibitem{teich}
B.Farb, D.Margalit, `` A primer on mapping class groups '', Princeton mathematical series.

\bibitem{belling}
P. Bellingeri, ``On presentation of surface braid groups'', arXiv:math/0110129 [math.GT]

\bibitem{BPZ}
A.A. Belavin, A.M. Polyakov, and A.B. Zamolodchikov, ``Infinite conformal symmetry in two-dimensional Quantum Field Theory'', Nucl. Phys. B241 (1984) 333-380

\end{thebibliography}
\end{document}